\documentclass[%
 reprint,
 superscriptaddress,
 amsmath,amssymb,
 aps,
 pre,
 onecolumn,
]{revtex4-2}
\usepackage{graphicx}
\usepackage{dcolumn}
\usepackage{bm}
\usepackage{upgreek}
\usepackage{bm}
\usepackage{amsmath}

\begin{document}

\preprint{APS/123-QED}

\title{Similarity solutions in elastohydrodynamic bouncing}

\author{Vincent Bertin}
\email[]{v.l.bertin@utwente.nl}
\affiliation{Physics of fluids group, University of Twente, 7500 AE Enschede, The Netherlands}

\date{\today}

\begin{abstract}
We investigate theoretically and numerically the impact of an elastic sphere on a rigid wall in a viscous fluid. Our focus is on the dynamics of the contact, employing the soft lubrication model in which the sphere is separated from the wall by a thin liquid film. In the limit of large sphere inertia, the sphere bounces and the dynamics is close to the Hertz theory. Remarkably, the film thickness separating the sphere from the wall exhibits non-trivial self-similar properties that vary during the spreading and retraction phases. Leveraging these self-similar properties, we establish the energy budget and predict the coefficient of restitution for the sphere. The general framework derived here opens many perspectives to study the lubrication film in impact problems.

\end{abstract}

\maketitle


\section{Introduction}

The impact or collision of a spherical object on a surface is a problem that has been of great interest for decades. Typical examples can be found at various length scales ranging from asteroid impact~\citep{chapman1994impacts}, drop impact~\citep{cheng2022drop}, ball sports~\citep{cross1999bounce}, collision in granular media~\citep{andreotti2013granular} or suspensions, or more commonly children playing with rubber bouncing balls. Of particular interest is the restitution coefficient, defined as the ratio between the bouncing speed $V_\infty$ and the impact speed $V_0$. An elastic collision corresponds to the case of a bouncing speed equal to the impact speed, without any loss of energy. In real impacts, various effects of the solid tend to decrease the restitution coefficient at large velocity~\citep{johnson1987contact} such as viscoelasticity~\cite{falcon1998behavior,ramirez1999coefficient}, plastic deformations~\cite{tabor1948simple,tsai1967study}, or sphere vibrations~\citep{hunter1957energy,reed1985energy,koller1987waves}. On top of these, viscous dissipation is predominant when the impact occurs in liquid environments, for instance in particle-laden flows~\citep{brandt2022particle}, which has a large variety of industrial and natural applications. Most numerical models of such flows use a point-particle approximation or immersed boundary and do not resolve the flow in the lubrication layer upon collision. Instead, empirical collision laws are implemented when particles start overlapping~\citep{brandle2013numerical,costa2015collision}. In this article, we aim at describing particle collisions, focusing on the canonical problem of a soft elastic sphere impacting a rigid surface in a viscous liquid. 

\begin{figure}
\centerline{\includegraphics{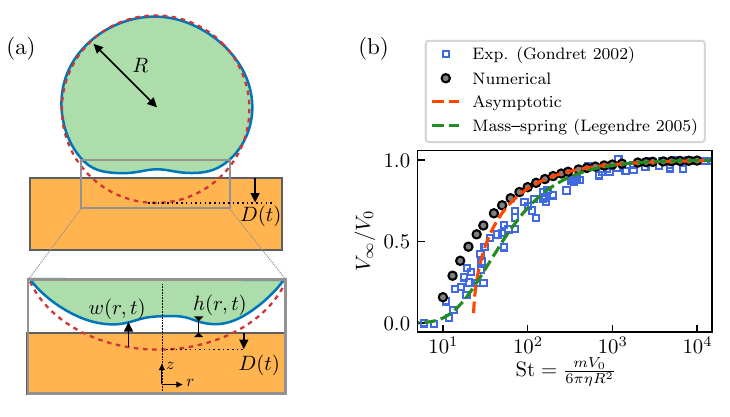}}	
\caption{\label{fig1} (a) Schematic of an elastohydrodynamic bouncing of a soft sphere on a rigid surface. The underformed sphere is indicated with red dashed lines. A thin film of thickness $h(r,t)$ prevents direct contact, where $r$ denotes the axisymmetric radial coordinate. The grey rectangle indicates a zoom in the contact region. (b) Bouncing velocity as a function of the Stokes number measured in the experiments of Gondret et al.~\citep{gondret2002bouncing}, in the numerical simulations and predicted by the asymptotic theory of Eq. \eqref{eq:restitution-coefficient}. The prediction of Legendre et al.~\citep{legendre2005experimental}, using a linear damped mass-spring model is shown in green dashed lines.}
\end{figure}

Several experimental studies have been devoted to the individual rebound of a particle on a surface, either immersed in a liquid~\cite{joseph2001particle,gondret2002bouncing}, or with a thin oil layer lubricating the surfaces~\cite{safa1986pressure,barnocky1988elastohydrodynamic,kaneta2007effects}. Head-on and oblique particle-particle collisions have also been investigated and follow a similar phenomenology~\cite{joseph2004oblique,yang2006dynamics}. The restitution coefficient increases with increasing velocity, as the contact time decreases and the viscous friction dissipates less energy (see Fig.~\ref{fig1}(b)). Remarkably, for a large variety of sphere sizes, material properties and fluid viscosities, the restitution coefficients collapse when plotted versus the Stokes number $\mathrm{St} = m V_0 / (6\pi\eta R^2)$, where $m$, $V_0$, $\eta$ and  $R$ are the sphere mass, impact velocity, viscosity and sphere radius. The Stokes number needs to exceed a critical number, which is of order of $10$, for bouncing to occur. Then, the restitution coefficient increases and reaches asympotically unity when the Stokes number is $\sim 1000$. An empirical expression $V_\infty/V_0 = \exp(-35/\mathrm{St})$ has been shown to describe fairly well the experimental data over the entire range of St~\citep{legendre2005experimental} (see green dashed lines in Fig.~\ref{fig1}(b)), and is motivated by an analogy with a damped spring-mass model. 

The bouncing dynamics is mainly modeled in two ways. First, reduced linear damped mass-spring models have been introduced~\citep{legendre2005experimental}. The non-linearities of the contact elasticity~\cite{schwager1998coefficient} and gravitational effects~\cite{falcon1998behavior} have also been taken into account. The second kind of models described the morphology of the lubrication layer upon contact. Davis et al. were the first ones to derive an elastohydrodynamic lubrication model, which describes the intricate coupling between the thin liquid film and elastic deformations, when particles move near soft interfaces~\cite{davis1986elastohydrodynamic}. The lubrication forces prevent the direct contact between the sphere and the rigid surface, such that a thin liquid film always separates the two surfaces. These models allow for a prediction of the critical bouncing Stokes number, that typically takes the form $\ln(H/\delta)$, where $H$ is the initial separation distance and $\delta$ is a cut-off length. The cut-off length is an elastohydrodynamic length for smooth surfaces, or the typical roughness for rough surfaces~\cite{barnocky1988elastohydrodynamic,yang2008mixed}. Piezoviscous and compressible effects, which play an important role in lubricants, have also been discussed~\cite{barnocky1989influence,wang2013central,venner2016central}. We also point out that the elastohydodynamic coupling gives rise to a very rich phenomenology for particle sliding and rotating near soft interfaces~\cite{salez2015elastohydrodynamics}.

The morphology of the lubrication layer during the contact has rarely been addressed~\cite{lian1996elastohydrodynamic}. The central film thickness has been identified to scale as $\mathrm{St}^{-1/2}$ in the Ref.~\cite{venner2016central}, suggesting the presence of self-similar solutions. We present in this article a detailed numerical and asymptotic analysis of the elastohydrodynamic bouncing. The Stokes number is set in the range $[10, 10^4]$ in the numerical simulations of this paper, which corresponds to the relevant experimental range (see Fig.~\ref{fig1}(b)). The structure of the lubrication layer during the impact dynamics is derived through self-similar solutions. Furthermore, the energy budget allows to find an asymptotic expression of the restitution coefficient at large Stokes number.

\section{Soft lubrication model}
\subsection{Formulation}
We briefly recall here the soft lubrication model, already employed to model the bouncing of sphere of radius $R$ and mass $m$ on a rigid planar surface~\cite{davis1986elastohydrodynamic,venner2016central,tan2019criterion}. We assume an impact normal to the surface such that the problem is axisymmetric, where the axis of symmetry is the sphere axis normal to the surface, referred later as vertical direction ($z$ in Fig.~\ref{fig1}(a)). We introduce the polar coordinates where the radial position is denoted $r$. The center of mass vertical position of the sphere, shifted by one radius, is denoted $D(t)$ (see Fig.~\ref{fig1} (a)). We suppose that the sphere is submerged in a viscous fluid of viscosity $\eta$. Due to lubrication forces, a thin film of liquid always separates the sphere from the surface and there is no direct contact. Using Newton's second law along the vertical direction, the sphere's dynamical equation is $m\dot{V}(t) = F(t)$, where $F(t)$ is the the vertical hydrodynamic viscous force acting on the particle, and $V(t) = \dot{D}(t)$ its vertical velocity. We focus here on the contact dynamics, such that the sphere is close enough to the surface to use the lubrication approximation. Therefore, the flow in the liquid gap is a parabolic Poiseuille flow and the liquid film thickness $h(r,t)$ follows the thin-film equation
\begin{equation}
	\label{eq:thin-film}
	\frac{\partial h(r,t)}{\partial t} = \frac{1}{12\eta r}\frac{\partial }{\partial r} \left(r h^3(r,t) \frac{\partial p(r,t)}{\partial r}\right),
\end{equation}
where $p(r,t)$ is the pressure field. Within the lubrication approximation, the contribution of the outer region of the contact to the hydrodynamic viscous force is negligible, and the hydrodynamic viscous force mainly comes from the contact region and follows the lubrication force as $F(t) = \int_0^\infty p(r,t) \, 2\pi r \mathrm{d}r$. Additionally, the sphere being very close to the surface, the underformed spherical particle can be modelled with a parabolic approximation such that the liquid gap can be written as~\cite{rallabandi2024fluid}
\begin{equation}
	\label{eq:film_thickness}
	h(r,t) = D(t) + \frac{r^2}{2R} - w(r,t),
\end{equation}
where $w(r,t)$ represents the elastic deformations (see Fig.~\ref{fig1} (a)). These deformations are modeled by using the linear elasticity theory and are related to the pressure field with a convolution integral involving the elastic Green's function. Integrating the convolution integral in the azimuthal direction leads to the integral relation~\citep{davis1986elastohydrodynamic}
\begin{equation}
	\label{eq:elasticity}
	w(r,t) = -\frac{4}{\pi E^*}\int_0^\infty \mathcal{M}(r,x) p(x,t) \, \mathrm{d}x, \quad \quad  \mathcal{M}(r,x) = \frac{x}{r+x}K\left(\frac{4rx}{(r+x)^2}\right), 
\end{equation}
where $E^* = E/(1-\nu^2)$ is the plane strain elastic modulus with $E$ and $\nu$ the Young's modulus and Poisson ratio respectively. The function $K$ is the complete elliptic integral of the first kind. 

We introduce the typical sphere elastic deformation length $D_0$ during the contact, 
\begin{equation}
\label{eq:hertz-indentation-scale}
D_0 = \left(\frac{mV_0^2}{E^*\sqrt{R}}\right)^{2/5},
\end{equation} 
obtained by balancing the kinetic energy $mV_0^2$ with the Hertz elastic energy $E^*\sqrt{R}D_0^{5/2}$ (see below). Equations~\eqref{eq:thin-film}-\eqref{eq:elasticity} are made dimensionless, by using the scales of the Hertz theory, respectively $D_0$, $\sqrt{RD_0}$, $D_0/V_0$, $p_0 = 2E^* \sqrt{D_0/R}/\pi$ and $F_0 = p_0 RD_0$ as vertical/radial length, time, pressure and force scales, as detailed in Appendix~\ref{app:dimensionless}. We then solved numerically the resulting equations using the finite-difference scheme proposed in Ref.~\citep{liu2022lubricated}, as discussed in Appendix~\ref{app:numerics}. As an initial condition, we suppose that the sphere's position is at a height $D_0$ with a downward velocity $V_0$. A single dimensionless number characterizes the bouncing dynamics, which is the Stokes number 
\begin{equation}
\label{eq:Stokes}
\mathrm{St} = \frac{mV_0}{6\pi \eta R^2}.
\end{equation}
The Stokes number can be interpreted as the ratio between sphere inertia ($m V_0^2$) to viscous dissipation ($6\pi \eta R^2V_0$), or alternatively the viscous dissipation time scale $mD_0/(6\pi\eta R^2)$ to the bouncing time scale $D_0/V_0$. For small Stokes number (large viscosity), or dissipation time smaller than the bouncing time, the sphere does not bounce as the entire initial kinetic energy is dissipated before contact. Therefore, in the context of bouncing, the Stokes number offers a quantification of the viscous dissipation during bouncing. The critical Stokes number for bouncing to occur depends slightly on the initial altitude and is typically of the order of $10$~\citep{davis1986elastohydrodynamic} (see Fig.~\ref{fig1} (b)). In this article, we focus on the intermediate to large Stokes number regime (low viscosity fluid), when the sphere bounces with a non-zero speed. We neglect the buoyancy forces here that do not influence on the contact dynamics as long as the work done by buoyancy during contact $\Delta \rho R^3 g D_0$ is small compare to the kinetic energy $mV_0^2$~\cite{falcon1998behavior}, where $\Delta \rho$ is the density difference between the sphere and the fluid. Lastly, we suppose that the impact speed is not too large, such that the typical sphere deformation $D_0$ is small compared to the sphere radius, corresponding to the condition $(mV_0^2/(E^* R^3))^{2/5} \ll 1$.

In the following sections, we will find various asymptotic similarity solutions for the film thickness, each with a specific scale different from the typical deformation $D_0$ (see Eq.~\eqref{eq:hertz-indentation-scale}). We shall notice that in the infinite Stokes limit, the elastohydrodynamic model converges toward the elastic collision, where the bouncing velocity is equal to the impact velocity, as discussed in the next subsection.

\section{Impact dynamics}

\subsection{Dry limit, $\mathrm{St}\to \infty$: Hertz theory}
\label{subsec:hertz}
In the absence of any surrounding fluid, or for $\mathrm{St}\to \infty$, an exact solution of the elastic bouncing dynamics has been introduced previously~\citep{johnson1987contact,hertz1881contact}. The vertical force acting on the sphere is zero as long as the sphere is out of contact, \textit{i.e.} $D(t)>0$. During the contact phase, the force follows the Hertz theory $4E^* R^{1/2} \delta^{3/2}(t)/3$, where we introduce $\delta(t) = -D(t)$ as the (positive) indentation. Substituting this expression in the Newton’s second law for the sphere, the indentation follows the ordinary differential equation
\begin{equation}
\label{eq:hertz-dynamics}
	m \ddot{\delta}(t) = -\frac{4E^*}{3} R^{1/2} \delta^{3/2}(t), \quad  \text{for }\delta(t) >0,
\end{equation}
which corresponds to a nonlinear mass-spring equation, where the spring stiffness increases with the indentation as $\sqrt{\delta}$. An exact solution of Eq. \eqref{eq:hertz-dynamics} can be expressed in the form of an implicit equation
\begin{equation}
\label{eq:hertz-dynamics-solution}
	\int_0^{\delta/D_0} \frac{\mathrm{d}x}{\sqrt{1-\frac{16}{15} x^{5/2}}} =  \, _2F_1\left(\frac{2}{5},\frac{1}{2};\frac{7}{5};\frac{16 (\delta(t)/D_0)^{5/2}}{15}\right) \frac{\delta(t)}{D_0} = \frac{V_0t}{D_0},
\end{equation}
where $_2F_1$ is the Hypergeometric function. We notice that the equivalent dynamical law for a linear spring model is $\arcsin(\delta(t)/D_0) = V_0t/D_0$, solution of Eq. \eqref{eq:hertz-dynamics-solution} if one replaces $16/15x^{5/2}$ by $x^2$ in the integral. The shape of Eq. \eqref{eq:hertz-dynamics-solution} is fairly similar to the usual sinusoidal law for harmonic oscillators. Furthermore, the Hertz theory predicts that the pressure and elastic deformations are piecewise-defined functions that read
\begin{subequations}
\begin{alignat}{3}
\label{eq:hertz-pressure}
	p_\mathrm{Hz}(r,t) &= \left\{\begin{array}{l}
		\frac{2E^*}{\pi R} \sqrt{a^2(t)-r^2}, \quad r<a(t), \\
		0, \quad r>a(t), \\
	\end{array}\right.\\
	w_\mathrm{Hz}(r,t) &= \left\{\begin{array}{l}
	-\delta(t) + \frac{r^2}{2R}, \quad r<a(t), \\
	- \frac{1}{\pi R}\left[(2a^2(t)-r^2)\arcsin\left(\frac{a(t)}{r}\right)+a(t)\sqrt{r^2-a^2(t)}\right], \quad r>a(t), \\
\end{array}\right.
\end{alignat}
\label{eq:hertz-field}%
\end{subequations}
where $a(t) = \sqrt{\delta(t) R}$ is the contact radius. The equations~\eqref{eq:hertz-field} are solutions of the elasticity equation (Eq.~\eqref{eq:elasticity}), satisfying the boundary conditions of the Hertz contact problem: (i) stress free condition outside the contact area $r>a$, and (ii) imposed elastic deformations in the contact zone $r<a$ with a parabolic shape. Injecting the Hertz elastic deformation in the film thickness definition Eq.~\eqref{eq:film_thickness}, we can express the film thickness predicted the Hertz theory
\begin{equation}
h_\mathrm{Hz}(r,t) = \left\{\begin{array}{l}
	0, \quad r<a(t), \\
	-\delta(t) + \frac{r^2}{2R} + \frac{1}{\pi R}\left[(2a^2(t)-r^2)\arcsin\left(\frac{a(t)}{r}\right)+a(t)\sqrt{r^2-a^2(t)}\right], \quad r>a(t), 
	\end{array}\right.
\end{equation}
that is indeed zero in within the contact zone $r<a(t)$. The limiting behavior of the Hertz pressure and film thickness in the vicinity of the contact radius is given by
\begin{equation}
\label{eq:hertz_thickness-singularity}
	h_\mathrm{Hz}(r\to a^+) = \frac{8\sqrt{2}\sqrt{a(t)}}{3\pi R} \left(r-a(t)\right)^{3/2}, \quad \quad 	p_\mathrm{Hz}(r\to a^-) = \frac{2E^*}{\pi R}\sqrt{2a(t)} \left(a(t)-r\right)^{1/2}.
\end{equation}
All these results will turn out relevant as ‘‘outer solutions" for the impact dynamics at finite Stokes number.

\subsection{Finite Stokes number}
\label{subsec:dynamics_finite_St}
\begin{figure}
\centerline{\includegraphics{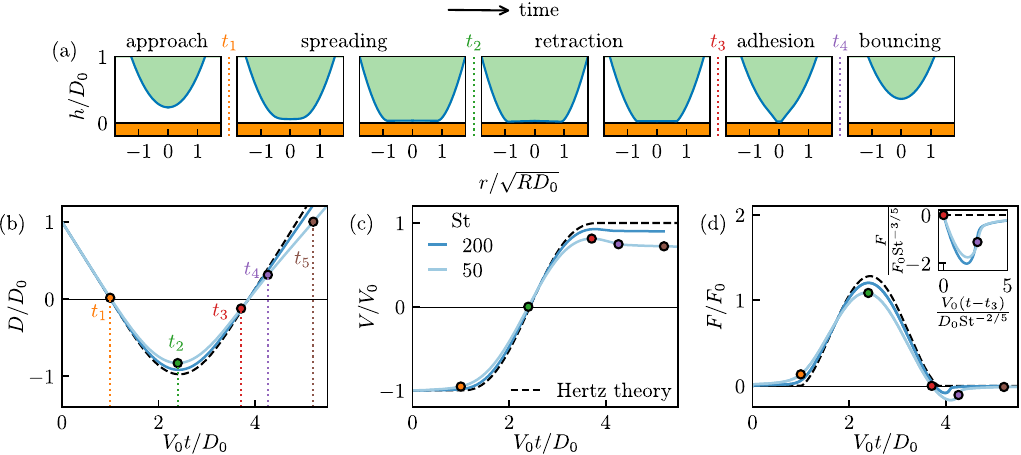}}	
\caption{\label{fig2} Bouncing dynamics: (a) Snapshots of the soft sphere interface during bouncing, which illustrates the different phases of the dynamics. The dimensionless times from left to right are $V_0 t/D_0 = 0.8, 1.2, 2.0, 2.8, 3.5, 4.1, 4.3$ and the Stokes number is $100$. Variation of the dimensionless sphere center of mass position (b), velocity (c) and lubrication force (d), versus the dimensionless time for two Stokes numbers $\mathrm{St} = 50$ and $200$. The black dashed lines indicate the Hertz theory \eqref{eq:hertz-dynamics-solution}, corresponding to the absence of viscous dissipation ($\mathrm{St} = \infty$). The points illustrate the five characteristic times separating the different phases for the case $\mathrm{St} = 50$ (see definition in the text). The inset in the panel (d) displays a zoom on the viscous adhesion phase, where both the horizontal and the vertical axis are rescaled with $\mathrm{St}^{-2/5}$ and $\mathrm{St}^{-3/5}$ respectively. We also point out that the origin of time has been shifted in the inset by $t_3$ where the force vanishes.}
\end{figure}

The bouncing dynamics at finite Stokes number is illustrated in Fig.~\ref{fig2} (a) (see also Supp. Video 1~\citep{Supp}) for $\mathrm{St} = 100$,  corresponding to the numerical solution of Eqs.~\eqref{eq:thin-film}-\eqref{eq:elasticity}. We identify five different stages during the bouncing, namely the approach, spreading, retraction, adhesion and bouncing phases. These are illustrated in Figure~\ref{fig2}(a), where we also introduce times $t_i$, for $i\in [1,5]$, that separate the phases. Initially, the sphere approaches the surface and the elastic deformation remains small, \textit{i.e.}, of order $\mathrm{St}^{-1}$. After the time of contact $t_1 = D_0/V_0$, where the sphere would have touched the surface in the absence of surrounding fluid, the elastic deformation gradually increases as if the sphere spreads on the surface. The sphere reaches its maximal deformation at time $t=t_2$ and its vertical velocity then changes sign, such that the contact radius now retracts. The time $t=t_3$ corresponds to the instant where the lubrication force vanishes, and is followed by a phase where a negative force is exerted on the sphere such that the sphere effectively adheres to the surface. Finally, at $t_4$, the sphere takes off and enters the bouncing phase, where the elastic deformations are small. We define empirically $t_4$ as the local inflexion point of the total energy of the system versus time curve, that turns out to be a good estimation of the take-off time. The fifth time $t_5$ is defined as the time at which the sphere reaches back its original altitude and bouncing velocity $V_\infty= V(t_5)$. 

The Figures~\ref{fig2}(b)-(d) quantify the global bouncing dynamics via the sphere vertical position, velocity and lubrication force for the cases $\mathrm{St} = 50$ and $200$. The specific times $t_i$, for $i\in [1,5]$, are indicated with circles only for the case $\mathrm{St} = 50$. The sphere dynamics is fairly close to the Hertz theory of Eq. \eqref{eq:hertz-dynamics-solution} (see black dashed lines in Fig.~\ref{fig2} (b)-(d)). The larger the Stokes number, the closer is the dynamics to Eq. \eqref{eq:hertz-dynamics-solution}, as the viscous effects are less pronounced. Nevertheless, an important difference is that the velocity after the bounce is smaller than the impact velocity, \textit{i.e.} $V_\infty/V_0 <1$, due to the presence of dissipation. Figure~\ref{fig1} (b) displays the coefficient of restitution $V_\infty/V_0$ versus the Stokes number. The numerical results of coefficient of restitution compares fairly well with the experimental results of Gondret et al.~\citep{gondret2002bouncing} obtained for a variety of spheres of different materials/sizes and in various liquids.

A remarkable feature occurs at the end of the contact, where the lubrication force becomes negative, indicating some effective adhesion. Indeed, the sphere appears to briefly ``stick"  to the surface (see adhesion phase in Fig.~\ref{fig2} (a)). We stress that there is no surface adhesion here, and the negative force arises from the viscous liquid, specifically due to the viscous resistance during the filling of the liquid gap. Hence, the adhesion force here cannot be described by JKR-like theory, but rather by the viscous adhesion that is also called Stefan adhesion~\citep{wang2020dynamic}. The adhesion becomes less pronounced at large Stokes number, highlighting the viscous nature of the force. Interestingly, an empirical collapse of the adhesion force is obtained when rescaling the force by $\mathrm{St}^{-3/5}$ and time by $\mathrm{St}^{-2/5}$ (see inset in Fig.~\ref{fig2} (d)), suggesting some universality of the viscous adhesion~\cite{wang2020dynamic}. The re-scaling leads to dimensional force and time scales respectively given by $\eta R V_0 \left[\eta V_0 / (R E^*)\right]^{-2/5}$ and $R/V_0 \left[\eta V_0 / (R E^*)\right]^{2/5}$, which are mass free and correspond to the usual elastohydrodynamic lubrication scales (see discussion in Appendix~\ref{app:scales}). A precise description of the flow and the elastohydrodynamic coupling in the adhesion phase is left for future work. 

The Hertz model is dissipation free, and thus cannot predict the coefficient of restitution. To understand the viscous dissipation during bouncing, we extend the Hertz theory and analyse the lubrication film thickness during the contact. We first focus on the spreading phase of the dynamics in section~\ref{sec:spreading}, and the retraction phase will be discussed in Section~\ref{sec:retraction}. The main assumption is that the Stokes number is large, so that the dynamics is close to Hertz theory. 

\section{Spreading phase}
\label{sec:spreading}

The film thickness and pressure profiles are shown in Fig.~\ref{fig3}(a) and (c) respectively for three different times during the spreading phase for the case of St=1000. The numerical pressure profiles agree very well with the solution of the Hertz theory Eq. \eqref{eq:hertz-field} within the contact region $r<a$. Hertz theory neglects the lubrication pressure. However, the pressure gradient $\mathrm{d}p/\mathrm{d}r$ is singular at the edge of the contact radius $r=a$ (see Eq. \eqref{eq:hertz_thickness-singularity}), so that at finite velocity, the film thickness must be small in order to balance Hertz and lubrication pressures. This gives rise to small zone near the contact radius, smoothing out the pressure gradient, as shown in Fig.~\ref{fig3}(c). Moreover, the liquid film profile presents a dimple shape, similar to drop impact~\cite{josserand2016drop}, or the profiles obtained during the thin-film drainage between droplets or bubbles~\cite{yiantsios1990buoyancy,chan2011film}. We separate the theoretical analysis of the film into two zones: namely the edge of the contact called “neck region" and the central region called “dimple region" (see schematics in Fig.~\ref{fig3}).

\subsection{Neck region: soft lubrication inlet analogy}
\label{subsec:neck-spreading}
\begin{figure}
\centerline{\includegraphics{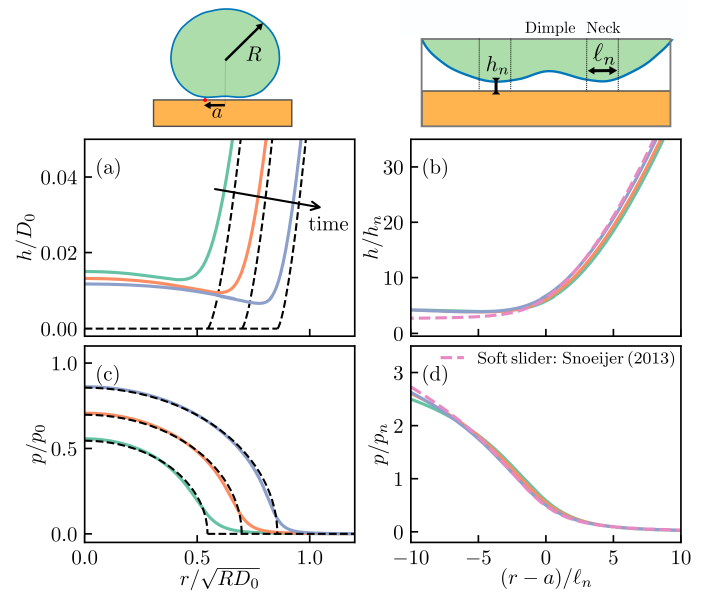}}	
\caption{\label{fig3} Spreading phase: neck solution at different times. Typical dimensionless film thickness (a) and pressure (c) as a function of the dimensionless radius for three different times (resp. $t = 1.3, \, 1.5$ and $1.8 \, D_0/V_0$) during the spreading phase. The Stokes number is set to $1000$. The black dashed lines show the Hertz theory. In (b) and (d), the profiles are rescaled by the typical length and pressure scales in the neck region, corresponding to Eq. \eqref{eq:similar-ansatz-neck}. The different lateral scales of the problem are shown in the schematic on top. The soft slider solution of Snoeijer et al.~\citep{snoeijer2013similarity} is shown in pink dashed lines.}
\end{figure}

The impact dynamics of a soft sphere exhibit resemblances with the problem of an elastic sphere sliding near a wall under an applied load~\cite{venner2016central}. Indeed, using as a reference frame $r\rightarrow r-a(t)$, where $a(t)$ is the contact radius of the Hertz theory, we can map the impact dynamics in the neck to the sliding of a soft sphere near a rigid wall, with a time-dependent velocity $u = \dot{a}$ and under a load $4E^* a^3 / (3R)$. In Refs.~\cite{bissett1989line,bissett1989lineII,snoeijer2013similarity}, it has been shown that the steady soft sliding has self similar solutions near the edge of the contact radius, that can be demonstrated rigorously using asymptotic matching methods. Here, we first recall the typical scales in the region using the scaling arguments~\citep{snoeijer2016analogies}. To do so, we introduce $h_n, \ell_n$ and $p_n$ as the typical film thickness, lateral length and pressure scales respectively in the neck region (see schematic in Fig.~\ref{fig3}). First, the fluid momentum balance, \textit{i.e.} Stokes equation $\eta\nabla^2\mathbf{u} = \nabla p$, in the horizontal direction yields to $\eta u/h^{2}_n = p_n/\ell_n$ in the lubrication limit. Then, following the linear elasticity theory, the pressure is proportional to the typical strain which reads $p_n = E^* h_n/\ell_n$. Finally, the similarity solution has to match the singular behavior of the Hertz theory near the contact radius (see Eq. \eqref{eq:hertz_thickness-singularity}). This imposes the geometric condition $h_n = a^{1/2} \ell_n^{3/2}/R$. Combining the aforementioned expressions, one finds 
\begin{equation}
\label{eq:spreading-neck-scale}
h_n = \frac{a^2}{R} \lambda^{3/5}, \quad \quad \ell_n = a \lambda^{2/5}, \quad \quad p_n = E^* \frac{a}{R} \lambda^{1/5}, \quad \quad \lambda = \frac{\eta u R^3}{E^*a^4}.
\end{equation}
The dimensionless group $\lambda$ is the relevant elastohydrodynamic dimensionless number, which needs to be small to enforce the hierarchy of scales $h_n \ll \ell_n \ll a \ll R$ of the asymptotic expansion. Following this approach, we introduce a self-similarity ansatz of the form
\begin{equation}
\label{eq:similar-ansatz-neck}
h(r,t) = h_n \mathcal{H}(\xi), \quad \quad p(r,t) = p_n\mathcal{P}(\xi), \quad \quad \xi = \frac{r-a}{\ell_n}.
\end{equation}
Substituting Eq. \eqref{eq:similar-ansatz-neck} into Eq. \eqref{eq:thin-film}, and using $\lambda \ll 1$, one gets
\begin{equation}
\label{eq:similar-neck-thin-film}
\left(\frac{\dot{h}_n \ell_n}{h_n \dot{a}}\mathcal{H} - \frac{\dot{\ell}_n}{\dot{a}}\xi \mathcal{H}' \right) - \mathcal{H}' = \frac{1}{12}\left(\mathcal{H}^3\mathcal{P}'\right)',
\end{equation}
where the term in bracket in the left hand side of Eq. \eqref{eq:similar-neck-thin-film} reflects the unsteadyness of the problem, given that both the velocity and load are time dependent. In the early times after contact, \textit{i.e.} when $t-t_1$ is small, the vertical velocity is roughly constant $V \approx -V_0$, such that the contact radius dynamics is governed by $a(t) \approx \sqrt{R V_0(t-t_1)}$ and $u(t) \approx \sqrt{R V_0/[4(t-t_1)]}$. Hence, the elastohydrodynamic parameter, which scales as $\lambda \propto (u/a^4)\propto (t-t_1)^{-5/2}$, diverges as $t\to t_1$. The steady similarity solution develops after a transient regime, once $\lambda \ll 1$. The typical scale of the transient time can be estimated as the one when $\lambda = 1$, leading to $(t-t_1) \sim (D_0/V_0) \mathrm{St}^{-2/5}$, which is much smaller than the bouncing time scale $D_0/V_0$ in the large $\mathrm{St}$ limit. We recover the elastohydrodynamic lubrication time scales, also governing the adhesion phase (see Appendix~\ref{app:scales}). Hence, after this very brief transient state, the unsteady terms of Eq. \eqref{eq:similar-neck-thin-film} are negligible and we can integrate to obtain 
\begin{equation}
\label{eq:similar-neck-thin-film-integrated}
\mathcal{P}' = 12 \frac{\mathcal{H}_0-\mathcal{H}}{\mathcal{H}^3},
\end{equation}
where $\mathcal{H}_0$ is an integration constant. In the limit where the length of the neck is smaller than the contact radius, \textit{i.e.} $\ell_n \ll a$, the elastic kernel $\mathcal{M}$ of Eq. \eqref{eq:elasticity} reduces to the dimensionless line force Green's function $ \mathcal{M}(r,r') \propto -\frac{1}{2} \ln\mid\xi-\xi'\mid$~\citep{johnson1987contact}. Taking (twice) the derivative of Eq. \eqref{eq:film_thickness}, and combining with the line force elastic kernel, we find
\begin{equation}
\label{eq:elasticity-similarity}
 \mathcal{H}''(\xi) = - \frac{2}{\pi} \int_\mathbb{R} \frac{\mathcal{P}'(\xi')}{\xi-\xi'} \, \mathrm{d}\xi',
\end{equation}
where we assume $\ell_n^2/(h_n R) \ll 1$, which follows from Eq. \eqref{eq:spreading-neck-scale}. The similarity solution has to match the asymptotic behavior of the Hertz theory near the contact radius (see Eq. \eqref{eq:hertz_thickness-singularity}). Writing the matching condition, we find
\begin{equation}
\label{eq:matching-condition}
\mathcal{H}(\xi \to \infty) = \frac{8\sqrt{2}}{2\pi} \xi^{3/2}, \quad \quad \mathcal{P}(\xi \to -\infty) = \frac{2\sqrt{2}}{\pi} (-\xi)^{1/2}.
\end{equation}
The equations \eqref{eq:similar-neck-thin-film-integrated}-\eqref{eq:matching-condition} are equivalent to the ones of the steady EHD sliding in the inlet zone of Snoeijer et al.~\cite{snoeijer2013similarity}, up to a trivial prefactor rescaling. As shown in Fig.~\ref{fig3}~(b) and (d), the rescaled profiles with the similarity variables of Eq. \eqref{eq:similar-ansatz-neck} indeed collapse for different times. Furthermore, the similarity solution derived in~\cite{snoeijer2013similarity} is in perfect agreement with the profile, demonstrating that the structure of the advancing neck in the bouncing problem maps perfectly to the inlet of the soft slider, in a quasi-steady fashion. It is not trivial that the advancing neck region in the bouncing problem is equivalent to the inlet of the steady soft sliding for two reasons. First, the bouncing problem is unsteady, and second it implies normal motion of the sphere versus lateral motion for the steady soft sliding. 

\begin{figure}
\centerline{\includegraphics{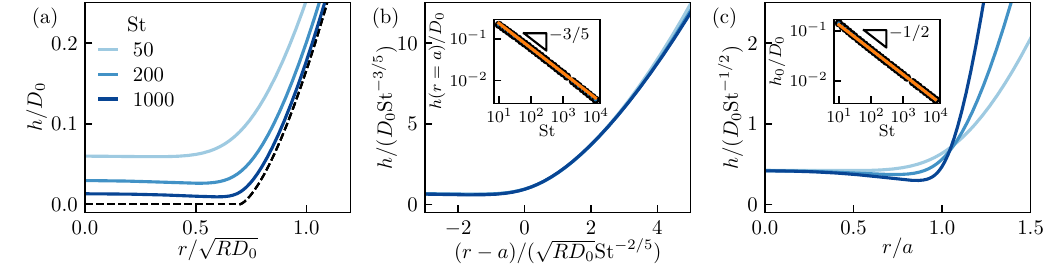}}	
\caption{\label{fig4} Spreading phase: Stokes number scaling. (a) Typical dimensionless film thickness as a function of the dimensionless radius at $t = 1.5 D_0/V_0$ during the spreading phase. The three colors indicate different Stokes number, respectively $50, \, 200, \, \text{and}\, 1000$ and the black dashed lines the Hertz theory. In (b) (resp. (c)), the thickness profiles are rescaled by the typical length scales in the neck (resp. dimple) region. The inset shows the thickness at the Hertz contact radius (b) and the central film thickness (c) versus the Stokes number in log-log, highlighting the Stokes number scaling with a fitted line. }
\end{figure}

The analysis above shows how at a fixed Stokes number, the spreading phase can be understood in time, by the analogy with a sliding contact problem. It is also of interest to investigate the scaling of the neck thickness upon varying the Stokes number -- in particular since we expect the neck to vanish in the limit $\mathrm{St} \to \infty$. In figure~\ref{fig4}(a), we thus plot profiles at different $\mathrm{St}$, at a fixed time ($t=1.5 D_0/V_0$) during the spreading phase. As expected, the larger the Stokes number, the thinner is the lubrication film and the closer it gets to the Hertz theory. The neck film thickness and lateral scales (see Eq. \eqref{eq:spreading-neck-scale}) can be rewritten to make explicit the Stokes number, as 
\begin{equation}
\frac{h_n}{D_0} =  (6\pi\mathrm{St})^{-3/5} \left(\frac{u D_0}{V_0\sqrt{RD_0}}\right)^{3/5} \left(\frac{a}{\sqrt{R D_0}}\right)^{-2/5}, \quad  \frac{\ell_n}{\sqrt{RD_0}} =  (6\pi\mathrm{St})^{-2/5} \left(\frac{u D_0}{V_0\sqrt{RD_0}}\right)^{2/5} \left(\frac{a}{\sqrt{R D_0}}\right)^{-3/5},
\end{equation}
Hence, we find that the Stokes number scaling of the film thickness and lateral scale of the neck during the spreading phase are $\mathrm{St}^{-3/5} $ and $\mathrm{St}^{-2/5} $ respectively. Rescaling the profiles of Fig.~\ref{fig4}(a) with the corresponding Stokes number scaling, a collapse of film thickness profiles is indeed observed in Fig.~\ref{fig4}(b) close to the neck region. Furthermore, the inset shows the film thickness at the radial position of the Hertz contact radius which is in perfect agreement with $\mathrm{St}^{-3/5}$ over the full $\mathrm{St}$ number range explored numerically.

\subsection{Central region: dimple height model}
\label{subsec:dimple-spreading}

\begin{figure}
\centerline{\includegraphics{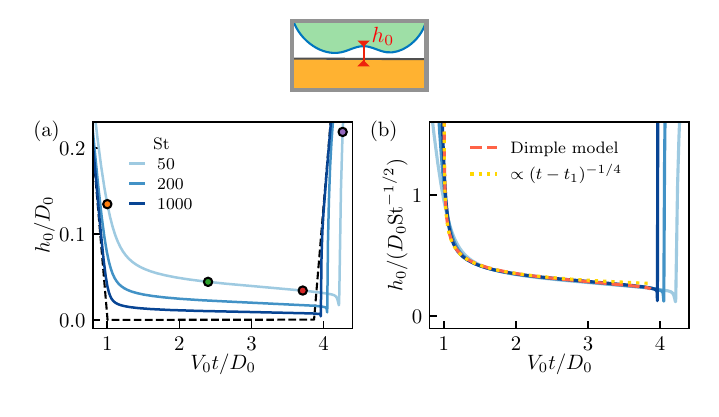}}	
\caption{\label{fig5} Dimple height. (a) Rescaled central film thickness height $h_0(t) = h(r=0,t)$ as a function of time in both the spreading and retraction phases. The colors indicate different Stokes numbers (same as in Fig.~\ref{fig4}), respectively $50$, $200$ and $1000$ from light to dark blue. As in Fig.~\ref{fig2}, the dots indicates the times separating the different phases of the bouncing dynamics. The panel (b) shows the same data, where the vertical axis is rescaled by $\mathrm{St}^{-1/2}$. The prediction of the dimple height Eq. \eqref{eq:dimple-height} is shown in red dashed lines, and the small time-to-contact asymptotic Eq.~\eqref{eq:dimple-height_scaling} is displayed in yellow green line. }
\end{figure}

Toward the dimple region, the neck similarity solutions deviates from the steady EHD inlet (see Fig.~\ref{fig3}~(b)), that has a uniform thickness. Instead, in the bouncing problem, the film thickness has some spatial variations in the central region, taking the form of a dimple. The central region is not described by the same scaling laws as the neck region, and a different analysis must be adopted. We focus on the central height of the film $h_0(t) = h(r \to 0,t)$, that is referred as dimple height in what follows. The Figure~\ref{fig5}(a) reports the dimple height as a function of time for various Stokes numbers. As discussed above, the larger the Stokes number and the smaller the film thickness, as the lubrication pressure gets smaller. Interestingly, the dimple heights for different Stokes numbers collapse when rescaled by $\mathrm{St}^{-1/2}$ (see Fig.~\ref{fig5}(b)). In Fig.~\ref{fig4}(c), we show the radial dependence of the film thickness profiles rescaled by $\mathrm{St}^{-1/2}$, where the profiles indeed collapse near the center. Nevertheless, the profiles deviate away from the symmetry axis. Therefore, the $D_0 \mathrm{St}^{-1/2}$ thickness scale is only valid in the vicinity of the symmetry axis and does not collapse the film thickness in the entire dimple.


We turn to an analysis of the dimple height. The pressure fields is well described by the Hertz profile Eq. \eqref{eq:hertz-pressure} (see Fig.~\ref{fig3}(c)). Injecting the Hertzian pressure field in the thin film equation Eq.~\eqref{eq:thin-film} and investigating the limit $r \to 0$, we obtain an ordinary differential equation for $h_0(t)$ as
\begin{equation}
\left.\frac{\partial h(r,t)}{\partial t}\right|_{r\to 0} = \frac{\mathrm{d}h_0(t)}{\mathrm{d}t} = \left. \frac{1}{12\eta r}\frac{\partial }{\partial r} \left(r h^3(r,t) \frac{\partial p_\mathrm{Hz}(r,t)}{\partial r}\right) \right|_{r\to 0} = - \frac{E^* }{3\pi \eta R }\, \frac{h_0^3(t)}{a(t)}.
\end{equation}
Solving this equation with the initial condition $h_0(t\to t_1^+) = \infty$, we get 
\begin{equation}
\label{eq:dimple-height}
h_0(t) = \left( \frac{2E^* }{3\pi \eta R } \int_{t_1}^t \frac{\mathrm{d}\hat{t}}{a(\hat{t})}\right)^{-1/2} = D_0 \mathrm{St}^{-1/2} \left( 4\int_{Vt_1/D_0}^{Vt/D_0} \frac{\mathrm{d}\hat{t}}{a(\hat{t})/\sqrt{RD_0}}\right)^{-1/2}.
\end{equation}
Not only we recover the $\mathrm{St}^{-1/2}$ scaling, but the dynamics of the dimple height is quantitatively described by Eq.~\eqref{eq:dimple-height} as shown in Fig.~\ref{fig5}(b). At small time-to-contact, i.e. $t-t_1 \ll D_0/V_0$, the Hertz contact radius follows $a(t) = \sqrt{RV_0(t-t_1)}$. In this limit, the dimple height expression simplifies to
\begin{equation}
\label{eq:dimple-height_scaling}
h_0(t) \simeq \left( \frac{4E^* }{3\pi \eta R^{3/2}V_0^{1/2} } (t-t_1)^{1/2}\right)^{-1/2} \propto (t-t_1)^{-1/4}.
\end{equation}
Interestingly, the small time-to-contact asymptotic expression provides an excellent description of the full expression, as shown with the yellow dotted lines in Fig.~\ref{fig5}(b). We point out that the $-1/2$ scaling in Stokes number of the dimple height has already been discussed by Venner et al.~\cite{venner2016central}.

\section{Retraction phase}
\label{sec:retraction}

\begin{figure}
\centerline{\includegraphics{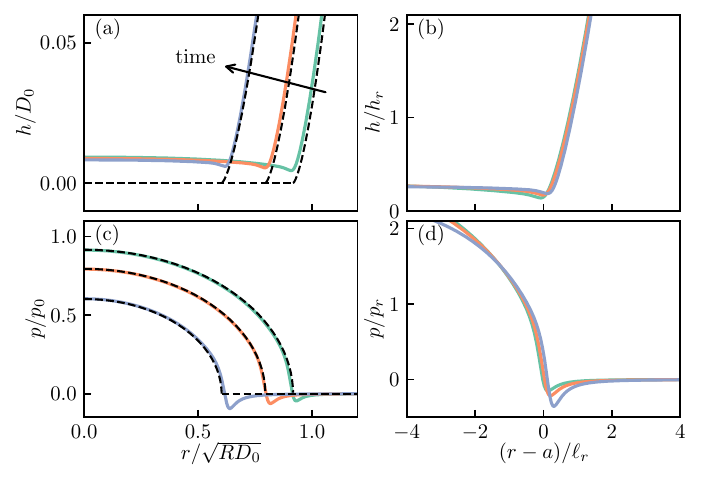}}	
\caption{\label{fig6} Retraction phase. Typical dimensionless film thickness (a) and pressure (c) as a function of the dimensionless radius for three different times (resp. $t = 2.9, \, 3.2$ and $3.5 \, D_0/V_0$) during the retraction phase, using the same notations as in Fig.~\ref{fig3}. In (b)-(d), the profiles are rescaled by the typical length and pressure scales in the neck during the retraction phase, corresponding to Eq.~\eqref{eq:scales-retraction}. In contrast to the spreading phase, there is no universal behavior, although the collapse is fairly good. }
\end{figure}

We now turn to the retraction phase of the bouncing dynamics, when the sphere vertical velocity is positive. 
Figure~\ref{fig6}(a) displays the film thickness and pressure profiles at various times during the retraction phase. The neck appears to be translated with minor changes of its vertical or lateral scales, which differs qualitatively from the spreading phase (see Fig.~\ref{fig3}(a)). The central region is fairly flat, as if the dimple has disappeared and the central thickness decreases very slowly in time, following Eq.~\eqref{eq:dimple-height}.


%

\begin{figure}
\centerline{\includegraphics{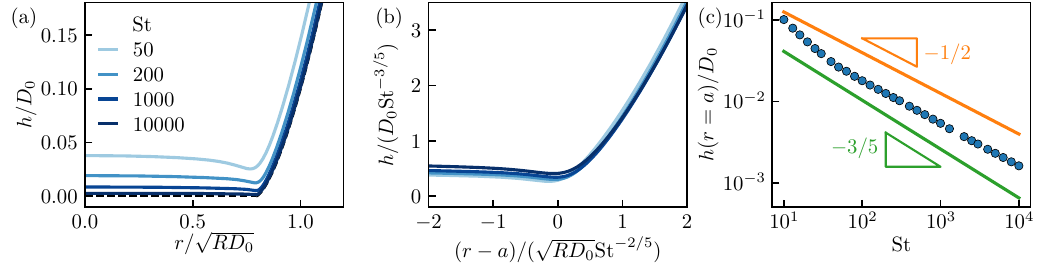}}	
\caption{\label{fig4_retraction} Retraction phase: Stokes number scaling. (a) Typical dimensionless film thickness as a function of the dimensionless radius at $t = 3.2 D_0/V_0$ during the spreading phase. The four colors indicate different Stokes number, respectively $50, \, 200, \, 1000\, \text{and}\, 10000$ and the black dashed lines the Hertz theory. In (b), the thickness profiles are rescaled by the typical length scales in the neck region during the spreading phase (see Fig.~\ref{fig4}). The panel (c) shows the thickness at the Hertz contact radius versus the Stokes number in log-log. The two lines indicates power laws with exponents $-3/5$ and $-1/2$ respectively.}
\end{figure}

Given the striking agreement between the neck region in the spreading phase with the soft slider inlet, we use the same approach to the retraction phase. The contact radius is now receding, such that the analogy must involve the outlet region of the soft slider. We stress that the central region of the soft slider has a uniform film thickness $\mathcal{H}_0 h_n$ that is universal and selected by the inlet profile, where $\mathcal{H}_0 =2.478...$ is a numerical prefactor~\citep{snoeijer2013similarity}. For the soft slider, the very same scaling arguments of section~\ref{subsec:neck-spreading} still apply in the outlet and a similarity solution can be found with the same scales as Eq. \eqref{eq:spreading-neck-scale}. For the impact, however, the situation is different: rescaling the numerical profiles in the neck with the scales Eq. \eqref{eq:spreading-neck-scale} does not collapse the data during the retraction phase. For instance, Fig.~\ref{fig4_retraction}(a) shows film thickness profiles at a fixed time $t = 3.2 D_0/V_0$ for various Stokes numbers. The data are rescaled in Fig.~\ref{fig4_retraction}(b) in the same manner as in the spreading phase (see Fig.~\ref{fig4}(b)), where we no longer observe a perfect collapse. The difference is further illustrated in Fig.~\ref{fig4_retraction}(c) where the value of the film thickness at the Hertz contact radius is displayed versus the Stokes number in log-log scale. In the range explored numerically, the film thickness does not follow the $\mathrm{St}^{-3/5}$ scaling as in the spreading phase, but is closer to the $\mathrm{St}^{-1/2}$ scaling of the dimple.

The fundamental difference between the inlet and outlet solution of the soft slider is that the inlet has a unique solution with a universal dimensionless flux $\mathcal{H}_0$, while the outlet solution has a family of solutions with varying fluxes, where the adopted self-similar shape is found by matching to the central region thickness~\citep{snoeijer2013similarity}. We note that such a qualitative feature is also found in the Bretherton bubbles in a tube~\citep{bretherton1961motion}, where the film thickness is selected from the universal solution in the front dynamical meniscus, while the rear adopts the same self-similarity but where the solution is non-unique and selected by matching the film thickness to the central solution. Hence, we conjecture that the typical thickness scale of the neck region in the retraction phase is selected by the matching to the central film thickness $h_0$, and not by the elastohydrodynamic film thickness $h_n$. As a result, the similarity solution is no longer universal and depends on a dimensionless number which is the instantaneous ratio between these two thicknesses $h_0/h_n$, that is actually large given the $\mathrm{St}$ scaling. For $\mathrm{St} = 10^3$, we observed numerically that the central film thickness in the retraction phase is roughly constant in time and scales as $h_0 \sim D_0 \mathrm{St}^{-1/2}$ (see Fig.~\ref{fig6}(a)). Hence, we suppose that the film thickness scale in the neck region during the retraction is $h_r = D_0 \mathrm{St}^{-1/2}$, where the subscript $r$ stands for retraction. As in section~\ref{subsec:neck-spreading}, we invoke the matching condition of the retracting neck solution to the singular behavior of Hertz theory (see Eq.~\eqref{eq:hertz_thickness-singularity}) to determine the lateral length and pressure scales in the neck as $h_r = \ell^{3/2}_r a^{1/2}/R$ and $p_r = E^* a^{1/2}\ell_r^{1/2}/R$. Combining the latter expressions, we find the scales
\begin{equation}
\label{eq:scales-retraction}
h_r = D_0 \mathrm{St}^{-1/2}, \quad \quad \ell_r = \sqrt{RD_0} \mathrm{St}^{-1/3} \left(\frac{a}{\sqrt{RD_0}}\right)^{-1/3}, \quad \quad p_r = p_0 \mathrm{St}^{-1/6} \left(\frac{a}{\sqrt{RD_0}}\right)^{1/3}.
\end{equation}
Once rescaled with Eq. \eqref{eq:scales-retraction}, the thickness and pressure fields in the neck collapse fairly well, as shown in Fig.~\ref{fig6}(b) and (d). 
Nevertheless, as expected, there is no clear universal self-similar solution, and some details of the profiles are not described by the rescaling. Further investigations would be necessary to determine the exact asymptotic structure of the solution at large Stokes number. Such an analysis would required Stokes number much larger than $10^5$ in order to distinguish between the two exponents. 

\section{Energy dissipation and coefficient of restitution}

\begin{figure}
\centerline{\includegraphics{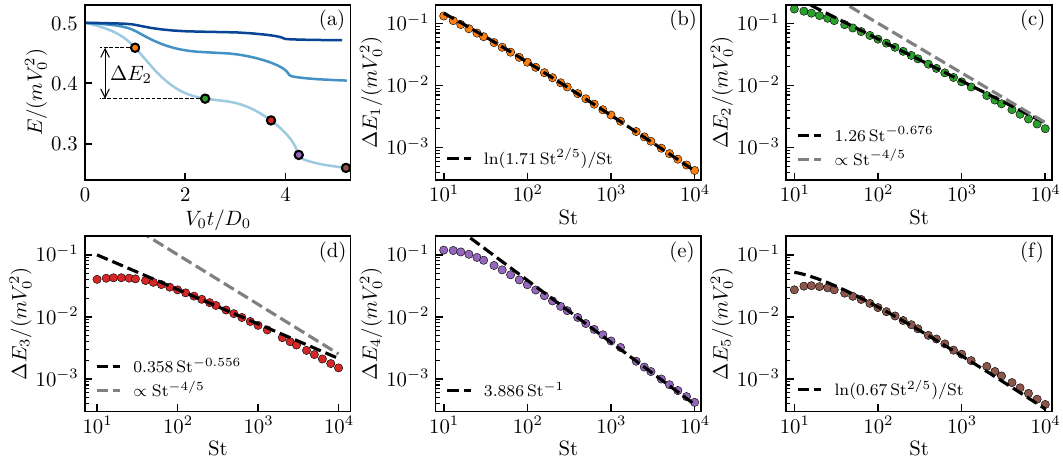}}	
\caption{\label{fig7} Energy budget. (a) Decay of the rescaled total energy of the sphere as a function of the dimensionless time. The colors indicate different Stokes numbers (same as in Fig.~\ref{fig4}), respectively $50$, $200$ and $1000$ from light to dark blue. The energy dissipated in each phase is denoted $\Delta E_i$, for $i \in [1, 5]$ and shown in (b)-(f) as a function of the Stokes number. The black dashed lines correspond to the fit of the numerical data with the asymptotic predictions in the large-Stokes limit. }
\end{figure}

Having discussed in details the evolution of the lubrication film during the contact dynamics, we now investigate the bouncing restitution coefficient. To predict the restitution coefficient, we analyse the energy budget during bouncing. The only source of dissipation comes from the liquid viscosity, such that the energy dissipation rate is given by 
\begin{equation}
\label{eq:dissipation-rate}
\frac{\mathrm{d}E}{\mathrm{d}t} = -\int_{\mathbb{R}^3} \eta (\nabla v)^2 \, \mathrm{d}^3x = - \int_{\mathbb{R}^2} \frac{h^3}{12\eta} (\nabla p )^2 \, \mathrm{d}^2x,
\end{equation}
where the total viscous dissipation in Eq.~\eqref{eq:dissipation-rate} is simplified to keep the dominant term within the lubrication approximation (see appendix C in Ref.~\cite{bertin2021capillary}). The total energy of the system, including kinetic energy and elastic energy of the sphere is denoted $E$ and is shown in Fig.~\ref{fig7}(a). Interestingly, the amount of energy dissipated in each phase of the dynamics is of the same order of magnitude, which motivates a model accounting for all the phases of the bouncing dynamics. We define the energy dissipated in each phase of the dynamics by $\Delta E_i = E(t_{i})-E(t_{i-1})$, for $i \in [1, 5]$ and where $t_0 = 0$, and aim at describing the asymptotic behaviour of each phase in the large $\mathrm{St}$ limit. 

First, during the approach phase, the elastic deformations are small (see Fig.~\ref{fig2}(a)), such that the lubrication force can be approximated by the lubrication force of rigid planar surface $F(t) = 6\pi\eta R^2 V_0 / D(t)$. Here we suppose that the velocity is asymptotically equal to the impact velocity $V(t) \simeq -V_0$, such that the distance decreases linearly in time, as $D(t) \simeq H - V_0 t$. Here, we generalize the result to an arbitrary initial height (denoted $H$), which is $D_0$ in the numerical simulations of Fig.~\ref{fig7}. Without elastic deformations, the energy dissipation rate can be expressed by $F(t)V(t) = -6\pi\eta R^2 V_0^2 / D(t)$. Integrating the dissipation rate, we obtain the amount of energy dissipated as
\begin{equation}
\label{eq:energy_drop15}
\Delta E_1 = 6\pi \eta R^2 V_0 \ln\left(\frac{H}{\delta}\right) = mV_0^2 \mathrm{St}^{-1} \ln\left(\frac{H}{\delta}\right),
\end{equation}
where one needs to introduce a cut-off length $\delta$, \textit{a priori} unknown, to regularize the integral. Similar expressions have already been derived previously~\citep{davis1986elastohydrodynamic}. As shown in Appendix~\ref{app:scales}, the typical film thickness scale at the transition between the approach and spreading phase is given by $D_0 \mathrm{St}^{-2/5}$, which is a good candidate of regularizing length. Hence, we fit the numerical energy drop in the approach phase with an asymptotic law $mV_0^2 \mathrm{St}^{-1} \ln\left(A_{1} \mathrm{St}^{2/5}\right)$, where $A_{1}$ is a fitting constant. An excellent agreement is found with the numerical data (see Fig.~\ref{fig7}(b)), where the fitting parameter is $A_1 = 1.71$. The same arguments can be employed for the bouncing phase, such that the energy drop should also follow the asymptotic law Eq. \eqref{eq:energy_drop15}, but with a different cut-off length. Fitting the numerical value of $\Delta E_5$ with $mV_0^2 \mathrm{St}^{-1} \ln\left(A_{5} \mathrm{St}^{2/5}\right)$ gives an excellent agreement (see Fig.~\ref{fig7}(f)), where $A_5=0.67$. 

In the spreading and retraction phase, the liquid flux is maximum in the neck region, such that the majority of the viscous dissipation occurs in this region. Estimating the viscous dissipation rate in Eq. \eqref{eq:dissipation-rate} as $h_{n,r}^3 (p_{n,r}/\ell_{n,r})^2 (a\ell_{n,r})/\eta$, and using the scales identified in sections \ref{subsec:neck-spreading} and \ref{sec:retraction}, the energy drop is asymptotically 
\begin{equation}
\label{eq:energy_asymptotic}
\frac{\Delta E_2}{mV_0^2} = A_2 \,\mathrm{St}^{-4/5}, \quad \quad \frac{\Delta E_3}{mV_0^2} = A_3\, \mathrm{St}^{-1/2},
\end{equation}
 where $A_{2,3}$ are numerical constants. In this case, the fitting of the numerical energy drop in the range of Stokes number explored numerically (i.e. $\mathrm{St} \in [10, 10^4]$) does not provide a very good fit. To get an approximate expression more accurate to describe the experimentally relevant range (see Fig.~\ref{fig1}(b)), we fit the numerical energy drop in the range $\mathrm{St} \in [10^2, 10^3]$, leaving the exponent as a fitting parameter. A very good agreement is found by using 
\begin{equation}
\label{eq:fitting_energy}
\frac{\Delta E_2}{mV_0^2} = 1.26\, \mathrm{St}^{-0.676}, \quad \quad \frac{\Delta E_3}{mV_0^2} = 0.358\, \mathrm{St}^{-0.556},
\end{equation}
as shown in Fig.~\ref{fig7}. The expression Eq.~\eqref{eq:fitting_energy} should not be seen as an exact asymptotic expression, but as a power-law fit in the Stokes range $\mathrm{St} \in [10^2, 10^3]$. Indeed, the energy drop would be well described by the exponent predicted by the elastohydrodynamic soft slider solution ($\mathrm{St}^{-4/5}$, see Eq.~\eqref{eq:energy_asymptotic}) only when the elastohydrodynamic parameter $\lambda^{1/5} \propto \mathrm{St}^{-1/5}$ (see Eq.~\eqref{eq:spreading-neck-scale}) is small. Such a condition is achieved for $\mathrm{St} \gg 10^5$, which is outside of the range explored numerically. Additional simulations at larger Stokes number would allow to determine the asymptotic energy drop during contact.  
 

Finally, the viscous adhesion phase has scales that are inertia-free, \textit{i.e.} independent of the mass of the sphere. Using the elastohydrodynamic scales (see Appendix~\ref{app:scales}), we expect an asymptotic energy drop of the form
\begin{equation} 
\label{eq:energy_drop_adhesion}
\frac{\Delta E_4}{mV_0^2} = A_4 \, \mathrm{St}^{-1}, 
\end{equation}
that is indeed mass-free. Fitting the numerical solutions of $\Delta E_4$ with \eqref{eq:energy_drop_adhesion} gives a very good agreement with $A_4 = 3.886$ as a fitting parameter. Then, the global energy budget reads $m (V_0^2-V_\infty^2)/2 = \sum_{i=1}^5 \Delta E_i$, which provides an expression for the restitution coefficient as
\begin{equation}
\label{eq:restitution-coefficient}
\frac{V_\infty}{V_0} = \sqrt{1 - 2\left[\mathrm{St}^{-1} \ln\left(1.71\,\mathrm{St}^{2/5}\right)  +1.26\, \mathrm{St}^{-0.676}+ 0.358\, \mathrm{St}^{-0.556},+ 3.886\,\mathrm{St}^{-1} + \mathrm{St}^{-1} \ln\left(0.67\,\mathrm{St}^{2/5}\right)\right]}.
\end{equation}
The asymptotic prediction of the restitution coefficient with the numerical data is excellent, as shown in Fig.~\ref{fig1}(b), which is not surprising as it relies on successive fitting of the energy drop through the dynamics. Most importantly, it provides an asymptotic expression of the restitution coefficient at large Stokes number, highlighting the physics at each phase of the bouncing. In the experimental range of Stokes number (see Fig.~\ref{fig1}(b)), the full expression is necessary to obtain a good approximation of the energy dissipated during the impact, as the energy drop in each phase is of similar magnitude (see Fig.~\ref{fig7}(a)). 

\section{Conclusion}

In this article, we have performed numerical simulations and asymptotic analysis of the elastohydrodynamic bouncing of a soft sphere on a rigid surface. We have demonstrated that the lubricated film thickness has non-trivial self-similar dynamics, that are analogous to the steady problem of a soft sliding sphere. Interestingly, the typical scales of the lubricated film are different in the spreading and retraction phases. The characterization of the self-similar features of the lubrication layer allows us to find an asymptotic expression of the restitution coefficient in elastohydrodynamic bouncing.

More generally, this article provides a general framework to study the coupling between the lubrication layer and interface deformations during impacts or collisions. Our model focuses on one of the most simple system, but many effects important in real impacts could be implemented, \textit{e.g.} surface roughness~\citep{wang2018morphology}, viscoelasticity~\citep{pandey2016lubrication}, adhesion~\citep{keh2016adhesion}, etc... An interesting extension of this work would be to consider oblique collisions~\citep{mindlin1953elastic,joseph2004oblique} and investigating the torque generated by the shear forces during the contact. Additionally, the present framework could be extended to a large variety of systems either changing the impactor, \textit{e.g.} drop impacts~\cite{mandre2009precursors}, elastic capsules~\cite{jambon2020deformation,remond2022dynamical}, or changing the impacted surface, \textit{e.g.}, stretched membranes~\cite{courbin2006impact,verzicco2021collision,aguero2022impact}, liquid pool~\cite{galeano2021capillary,sykes2023droplet,alventosa2023inertio} and so on. 

\bigskip
\textbf{Supplementary movie.}
A supplementary movie is available at 

\bigskip
\textbf{Acknowledgements.}
It is a pleasure to thank P. Chantelot, J. Eggers, A. Oratis and  J. H. Snoeijer for stimulating discussions, as well as D. Bonn, B. Gorin and N. Ribe for sharing unpublished experimental results that motivate this work. I also thank the anonymous referees for constructive feedback that greatly improved this paper.

\bigskip
\textbf{Founding.}

We acknowledge financial support from NWO through the VICI Grant No. 680-47-632. 

\bigskip
\textbf{Declaration of interests.}

The author reports no conflict of interest.

\bigskip
\textbf{Data availability statement.} 
The code developed in the present article is permanently available on GitHub~\citep{Git}. 

\bigskip
\textbf{Author ORCIDs.}

Vincent Bertin: https://orcid.org/0000-0002-3139-8846

\appendix

\section{Dimensionless equations}
\label{app:dimensionless}
The appendix expands the non-dimensionalization of the soft lubrication equations~\eqref{eq:thin-film}-\eqref{eq:elasticity}. We choose to use the typical elastic scales to make the equations dimensionless, which does not account for viscous effects. The typical elastic deformation of a sphere with an impact speed $V_0$ follows
\begin{equation}
D_0 = \left(\frac{mV_0^2}{E^*\sqrt{R}}\right)^{2/5}, 
\end{equation}
obtained by balancing the kinetic energy $mV_0^2$ with the elastic energy $E^* \sqrt{R}D_0^{5/2}$. Using the Hertz pressure profile, we introduce a pressure scale $p_0$ based on the typical elastic deformation as 
\begin{equation}
p_0 = \frac{2E^*}{\pi}\sqrt{\frac{D_0}{R}} = \frac{2E^*}{\pi} \left(\frac{mV_0^2}{E^*R^3}\right)^{1/5}.
\end{equation}
Hence, we introduce dimensionless variables with tilde as 
\begin{equation}
\label{eq:dimensionless}
\begin{split} 
h(r,t) = D_0 \tilde{h}(\tilde{r},&\tilde{t}), \quad \quad w(r,t) = D_0 \tilde{w}(\tilde{r},\tilde{t}), \quad \quad p(r,t) = p_0 \tilde{p}(\tilde{r},\tilde{t}), \quad \quad F(t) = p_0 RD_0 \tilde{F}(\tilde{t}), \\
D(t) &= D_0 \tilde{D}(\tilde{t}) \quad \quad V(t) = V_0 \tilde{V}(\tilde{t}) \quad \quad r = \sqrt{RD_0}\tilde{r}, \quad \quad t = \frac{D_0}{V_0} \tilde{t}.
\end{split}
\end{equation}
Substituting Eq. \eqref{eq:dimensionless} into \eqref{eq:thin-film}-\eqref{eq:elasticity}, we obtain the system of equations
\begin{subequations}
\label{eq:dimensionless-soft-lubrication}
\begin{equation}
\label{eq:dimensionless-thin-film}
	\frac{\partial \tilde{h}(\tilde{r},\tilde{t})}{\partial \tilde{t}} = \frac{\mathrm{St}}{\tilde{r}}\frac{\partial }{\partial \tilde{r}} \left(\tilde{r} \tilde{h}^3(\tilde{r},\tilde{t})  \frac{\partial \tilde{p}(\tilde{r},\tilde{t}) }{\partial \tilde{r}}\right),
\end{equation}
\begin{equation}
\label{eq:dimensionless-film-thickness}
	\tilde{h}(\tilde{r},\tilde{t}) = \tilde{D}(\tilde{t}) + \frac{\tilde{r}^2}{2} - \tilde{w}(\tilde{r},\tilde{t}),
\end{equation}
\begin{equation}
\label{eq:dimensionless-elasticity}
	\tilde{w}(\tilde{r},\tilde{t}) = -\frac{8}{\pi^2}\int_0^\infty \mathcal{M}(\tilde{r},\tilde{x}) \tilde{p}(\tilde{x},\tilde{t}) \, \mathrm{d}\tilde{x}, \quad \quad  \mathcal{M}(\tilde{r},\tilde{x}) = \frac{\tilde{x}}{\tilde{r}+\tilde{x}}K\left(\frac{4\tilde{r}\tilde{x}}{(\tilde{r}+\tilde{x})^2}\right), 
\end{equation}
\begin{equation}
\frac{\mathrm{d}\tilde{V}}{\mathrm{d}\tilde{t}}= \tilde{F}(\tilde{t}), \quad \quad \tilde{F}(\tilde{t}) = 4\int_0^\infty \tilde{p}(\tilde{r},\tilde{t}) \,  \tilde{r} \mathrm{d}\tilde{r},
\end{equation}
\begin{equation}
\tilde{V} = \frac{\mathrm{d}\tilde{D}}{\mathrm{d}\tilde{t}}.
\end{equation}
\end{subequations}
The sphere is assumed to start with an impact velocity $V_0$ and at a distance $D_0$ for the surface, leading to the initial conditions 
\begin{equation}
\tilde{D}(0) = 1, \quad \quad \tilde{V}(0) = -1.
\end{equation}
Initial conditions are also required for the pressure and deformation fields. A naive choice would be to set the initial pressure and deformation to zero, which leads to unphysical jumps at $t=0^+$ that affect the approach phase. To bypass this issue, we introduce an initialization phase where the velocity is ramped from $0$ to $-V_0$ during a time $t_\mathrm{ini}$, as suggested in Ref.~\cite{liu2022lubricated}. More precisely, the velocity and sphere position are set to
\begin{equation}
\tilde{V}(\tilde{t}) = -\left( 1 + \frac{\tilde{t}}{\tilde{t}_\mathrm{ini}}\right), \quad \quad \tilde{D}(\tilde{t}) = 1 -\left( \tilde{t} + \frac{\tilde{t}^2}{2\tilde{t}_\mathrm{ini}}\right), \quad \quad -\tilde{t}_\mathrm{ini} \leq \tilde{t} < 0,
\end{equation}
and the deformation and pressure fields are identically zero at $\tilde{t}=-\tilde{t}_\mathrm{ini}$. Then, solving the initialization phase with the numerical scheme discussed in Appendix \ref{app:numerics} allows to obtain initial pressure and deformation fields $\tilde{p}(\tilde{r},0)$ and $\tilde{w}(\tilde{r},0)$. We have checked that the choice of the initialization time has negligible effects in the overall bouncing dynamics. 

\section{Finite-difference scheme}
\label{app:numerics}
In this appendix, we detail the finite-difference scheme used to solve the dimensionless lubrication equations \eqref{eq:dimensionless-soft-lubrication}. We follow the methodology introduced by Liu et al.~\cite{liu2022lubricated}. We introduce a uniform spatial grid $\tilde{r}_i = i \,\Delta r$, for $i \in [0, N-1]$, where $\Delta r$ is the grid size and $N$ the number of radial points. The temporal axis is also discretized by using a constant time step $\Delta t$ as $\tilde{t}^n = n \Delta t$. The time step is set to $10^{-4}$ or below, which is small enough to avoid any negative values of $h$ and which ensures numerical stability. Hence, the pressure, film thickness and deformation fields are discretized as $\tilde{p}(\tilde{r},\tilde{t}) = \tilde{p}_i^n$, $\tilde{h}(\tilde{r},\tilde{t}) = \tilde{h}_i^n$ and $\tilde{w}(\tilde{r},\tilde{t}) = \tilde{w}_i^n$ and the velocity and sphere position $\tilde{V} = \tilde{V}^n$ and $\tilde{D} = \tilde{D}^n$ respectively. The dimensionless thin-film equation \eqref{eq:dimensionless-thin-film} can be expanded as 
\begin{equation}
\label{eq:dimensionless-thin-film_expanded}
\tilde{V} - \frac{\partial \tilde{w}(\tilde{r},\tilde{t})}{\partial t}   = \mathrm{St} \left[ \frac{\tilde{h}^3}{\tilde{r}} \frac{\partial \tilde{p}}{\partial \tilde{r}} +3\tilde{h}^2 \left(r - \frac{\partial \tilde{w}}{\partial \tilde{r}}\right) \frac{\partial \tilde{p}}{\partial \tilde{r}} + \tilde{h}^3\frac{\partial^2 \tilde{p}}{\partial \tilde{r}^2}\right],
\end{equation}
where we used the film thickness definition Eq. \eqref{eq:dimensionless-film-thickness}. The non linear term $h^3$ of the thin-film equation is evaluated at the previous time, so that we end up with a linear discrete set of equations, which greatly reduces the computational time as compared to non linear schemes. Furthermore, we use an implicit scheme which provides a better numerical stability as compare to explicit methods. Hence, Eq. \eqref{eq:dimensionless-thin-film_expanded} is discretized as  
\begin{equation}
\label{eq:dimensionless-thin-film_discrete}
\tilde{V}^{n+1} - \frac{\tilde{w}^{n+1}_i - \tilde{w}_i^n}{\Delta t} = \mathrm{St} \left[ \frac{(\tilde{h}^{n}_i)^3}{\tilde{r}_i} \left.\frac{\partial \tilde{p}}{\partial \tilde{r}}\right|^{n+1}_i +3(\tilde{h}^{n}_i)^2 \left(r_i - \left.\frac{\partial \tilde{w}}{\partial \tilde{r}}\right|^{n}_i\right) \left.\frac{\partial \tilde{p}}{\partial \tilde{r}}\right|^{n+1}_i + (\tilde{h}^{n}_i)^3\left.\frac{\partial^2 \tilde{p}}{\partial \tilde{r}^2}\right|^{n+1}_i\right],
\end{equation}
where the film thickness discretization is $\tilde{h}^n_i = \tilde{D}^n + \tilde{r}_i^2/2 - \tilde{w}_i^n$. The first and second order discrete spatial derivative of Eq. \eqref{eq:dimensionless-thin-film_discrete} are evaluated as 
\begin{equation}
\left.\frac{\partial \tilde{w}}{\partial \tilde{r}}\right|^{n}_i = \frac{\tilde{w}_{i+1}^n-\tilde{w}_{i}^n}{\Delta r}, \quad \quad \left.\frac{\partial \tilde{p}}{\partial \tilde{r}}\right|^{n+1}_i  = \frac{\tilde{p}_{i+1}^{n+1}-\tilde{p}_{i}^{n+1}}{\Delta r}, \quad \quad \left.\frac{\partial^2 \tilde{p}}{\partial \tilde{r}^2}\right|^{n+1}_i =  \frac{\tilde{p}_{i+1}^{n+1}-2\tilde{p}_{i}^{n+1}+\tilde{p}_{i-1}^{n+1}}{\Delta r^2}.
\end{equation}
The thin-film equation is a second-order differential equation in the radial coordinate, so that we need to introduce two boundary conditions. The symmetry condition imposes that the gradient of the pressure in the center is null, \textit{i.e.} $\frac{\partial \tilde{p}}{\partial \tilde{r}}(\tilde{r}=0)=0$. Furthermore, at large radius, the pressure field decays rapidly as $r^{-4}$ from Eq.  \eqref{eq:dimensionless-thin-film}. Hence, we impose the boundary conditions
\begin{equation}
p_1^{n} - p_0^n = 0, \quad \quad p_{N-1}^n=0.
\end{equation}
To evaluate the integral equation of elasticity \eqref{eq:dimensionless-elasticity}, we suppose that the pressure field is uniform and equal to $\tilde{p}^n_i$ on each domain $[r_i - \frac{\Delta r}{2}, r_i + \frac{\Delta r}{2}]$, for $i\geq 1$, and $p_0^n$ on the domain $[0,\frac{\Delta r}{2}]$ near the symmetry axis. This leads to the equation
\begin{equation}
\label{eq:discrete-elasticity}
\tilde{w}_i^n = -\frac{8}{\pi^2} \left[ p_0^n \int_0^{\frac{\Delta r}{2}} \frac{x}{x+\tilde{r}_i} K\left(\frac{4 x \tilde{r}_i}{(x+\tilde{r}_i)^2}\right)\mathrm{d}x  + \sum_{i=1}^{N-1} p_i^n\int_{\tilde{r}_i-\frac{\Delta r}{2}}^{\tilde{r}_i+\frac{\Delta r}{2}} \frac{x}{x+\tilde{r}_i} K\left(\frac{4 x \tilde{r}_i}{(x+\tilde{r}_i)^2}\right)\mathrm{d}x \right].
\end{equation}
The integrals in \eqref{eq:discrete-elasticity} are independent of the discrete fields, and only depends on the spatial grid. Therefore, they can be computed once and stored in a matrix to save some computational time. Numerically, these are evaluated with a Gaussian quadrature using the scipy integrate library in Python. Finally, the Newton’s second law for the sphere is discretized with a backward Euler scheme as
\begin{equation}
\frac{\tilde{V}^{n+1}-\tilde{V}^{n}}{\Delta t} = 4 \sum_{i=0}^{N-1} \tilde{p}_i^{n+1} \tilde{r}_{i} \Delta r, \quad \quad  \frac{\tilde{D}^{n+1}-\tilde{D}^{n}}{\Delta t} = \tilde{V}^{n+1},
\end{equation}
which complete the discrete system of equations. The code is available at the link~\citep{Git}.

\section{Elastohydrodynamic lubrication scales in the adhesion phase and the transition from approach to spreading.}
\label{app:scales}
\begin{figure}[t]
\centerline{\includegraphics{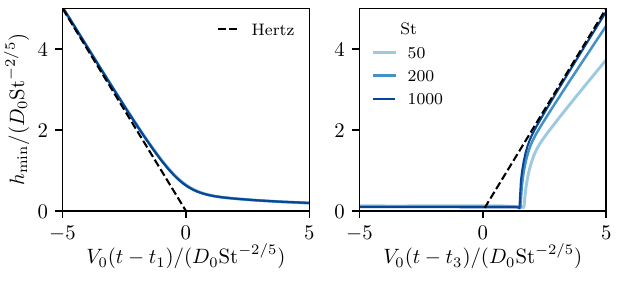}}	
\caption{\label{fig8} Dimensionless minimum film thickness versus the dimensionless time. On the left (resp. right) panel, the time is shifted by $t_1$ (resp. $t_3$). The film thickness and time axis are rescaled by $\mathrm{St}^{-2/5}$ corresponding to the mass-free elastohydrodynamic lubrication scales. }
\end{figure}

In this section, we rationalize the typical scales found in the main text for both the adhesion phase (see section~\ref{subsec:dynamics_finite_St}, and inset in Fig.~\ref{fig2}(d)) and the transition between approach to the spreading phase (see section~\ref{subsec:neck-spreading}). The corresponding scales are denoted below with star in superscript. Both of these phases occur on time scales much smaller than the bouncing time $D_0/V_0$. Hence, the kinetic energy of the sphere does not vary significantly during these phases and the scales must be independent of the sphere mass. The elasticity plays a role in the approach when the elastic deformations $w^*$ are comparable to the film thickness $h^*$, i.e. $h^*\sim w^*$. The film thickness scale can be found by balancing the typical lubrication pressure scale $\eta V_0 R/h^{*2}$ with the elastic pressure $E^* w^*/r^*$, where $r^*$ is the typical radial scales. The spherical shape of the film thickness in the contact region enforces the geometric relation $h^* = r^{*2}/R$. Combining these relationships, we recover the usual elastohydrodynamic lubrication scales as
\begin{equation}
\label{eq:scales_EHD}
w^* = h^* = R \left(\frac{\eta V_0}{E^*R}\right)^{3/5} = D_0 \mathrm{St}^{-2/5}, \quad \quad t^* = \frac{h^*}{V_0} = \frac{D_0}{V_0} \mathrm{St}^{-2/5}, \quad \quad F^* = \frac{\eta R^2 V_0}{h^*} = F_0 \mathrm{St}^{-3/5}.
\end{equation}
The figure~\ref{fig8} displays the minimum film thickness versus time, for times near the transition from approach to spreading (left) and in the adhesion (right) phase. An excellent collapse of the numerical profiles is found using $h^*$ and $t^*$ as scale which confirm that the elastohydrodynamic lubrication scales govern the corresponding processes.

\providecommand{\noopsort}[1]{}\providecommand{\singleletter}[1]{#1}%

\end{document}